\documentclass{iaus}
\usepackage{graphicx}
\usepackage{subfigure}

\title[~~La Silla-QUEST Variability Survey] 
{La Silla-QUEST Variability Survey \\ in the Southern Hemisphere}

\author[Hadjiyska et al.]   
{Ellie Hadjiyska$^1$,
 David Rabinowitz$^1$, Charles Baltay$^1$, \\ Nancy Ellman$^1$, Peter Nugent$^2$, Robert Zinn$^3$, Benjamin Horowitz$^1$, Ryan McKinnon$^1$, Lissa R. Miller$^3$}

\affiliation{$^1$Center for Astronomy $\&$ Astrophysics, Yale University, \\ P.O. Box 208120, New Haven, CT 06520-8120 USA \\ email: {\tt ellie.hadjiyska@yale.edu}\\[\affilskip] $^2$Lawrence Berkeley National Laboratory
M.S. 50B-4206 - 1 Cyclotron Road - Berkeley, CA, 94720-8139 USA\\[\affilskip] $^3$Dept. of Astronomy, Yale University, \\ P.O. Box 208101, New Haven, CT 06520-8101 USA}

\pubyear{2011}
\volume{285}  
\pagerange{???--???}
\jname{New Horizons in Time Domain Astronomy}
\editors{R.E.M. Griffin, R.J. Hanisch \& R. Seaman, eds.}
\begin{document}

\maketitle

\begin{abstract}
We describe the La Silla-QUEST (LSQ) Variability Survey. LSQ is a dedicated wide field synoptic survey in the Southern Hemisphere, focusing on the discovery and study of transients ranging from low redshift (z $<$ 0.1) SN Ia, Tidal Disruption events, RR Lyrae variables, CVs, Quasars, TNOs and others. The survey utilizes the 1.0-m Schmidt Telescope of the European Southern Observatory at La Silla, Chile with the large area QUEST camera, a mosaic of 112 CCD's with field of view of 9.6 square degrees. The LSQ Survey was commissioned in 2009, and is now regularly covering ~1000 square deg per night with a repeat cadence of hours to days. The data are currently processed on a daily basis. We present here a first look at the photometric capabilities of LSQ and we discuss some of the most interesting recent transient detections.
\end{abstract}

\firstsection 
\section{Introduction}

\begin{figure}[h]
\begin{center}
 \includegraphics[width=2.2in]{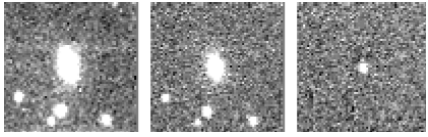}
 \caption{A possible supernova. From left -- reference image, night1, subtracted image.)}
   \label{fig3}
\end{center}
\end{figure}

After the completion of the Palomar-QUEST northern sky survey in September 2008 the QUEST Large Field Camera (\cite[Baltay et al. 2007]{Baltay_etal07})
was moved and installed on the 1.0-m ESO Schmidt in La Silla and had first light on April 24, 2009. Since September 2009, the southern survey has been
in routine observations (\cite[Andrews et al. 2008]{Andrews_etal08}) and the telescope and camera are controlled from Yale and fully robotic. We have 90$\%$ of the time
on the telescope with 10$\%$ allocated to Chile.
The QUEST camera consists of 112 CCDs of 600 $\times$ 2400 Sarnoff thinned pixels, back illuminated devices with 13 $\mu$m x 13 $\mu$m pixel pitch.
The camera covers an area of 4.6$^\circ$ $\times$ 3.6$^\circ$ on the sky and a plate scale of 0.86 arcsec/pixel. The survey covers $\sim$1000 square degrees per night, primarily between $\pm$25$^\circ$ to allow for follow up from both hemispheres.
The LSQ variability (SN and transient) survey uses 60 sec exposures (and the TNO survey-180 sec) taken twice a night with a cadence of 2 nights in one broad band filter of 4000 to 7000{\AA} (Qst*-band). The seeing at La Silla for the 60 sec exposures is 1.7 arcsec FWHM, reaching depth of 20.5 mag. The LSQ survey subtraction pipeline has started producing between 400 and 900 transient candidates each night (Fig.\,\ref{fig3}).




\section{Transient Detections}
\subsection{RR Lyrae Variables}

\begin{figure}[h]
 \begin{center}
   \begin{tabular}{c}
     \subfigure[RR Lyrae V magnitude light curves for RRab 12574, 10770 and 5381.]{\includegraphics[width=2.3in]{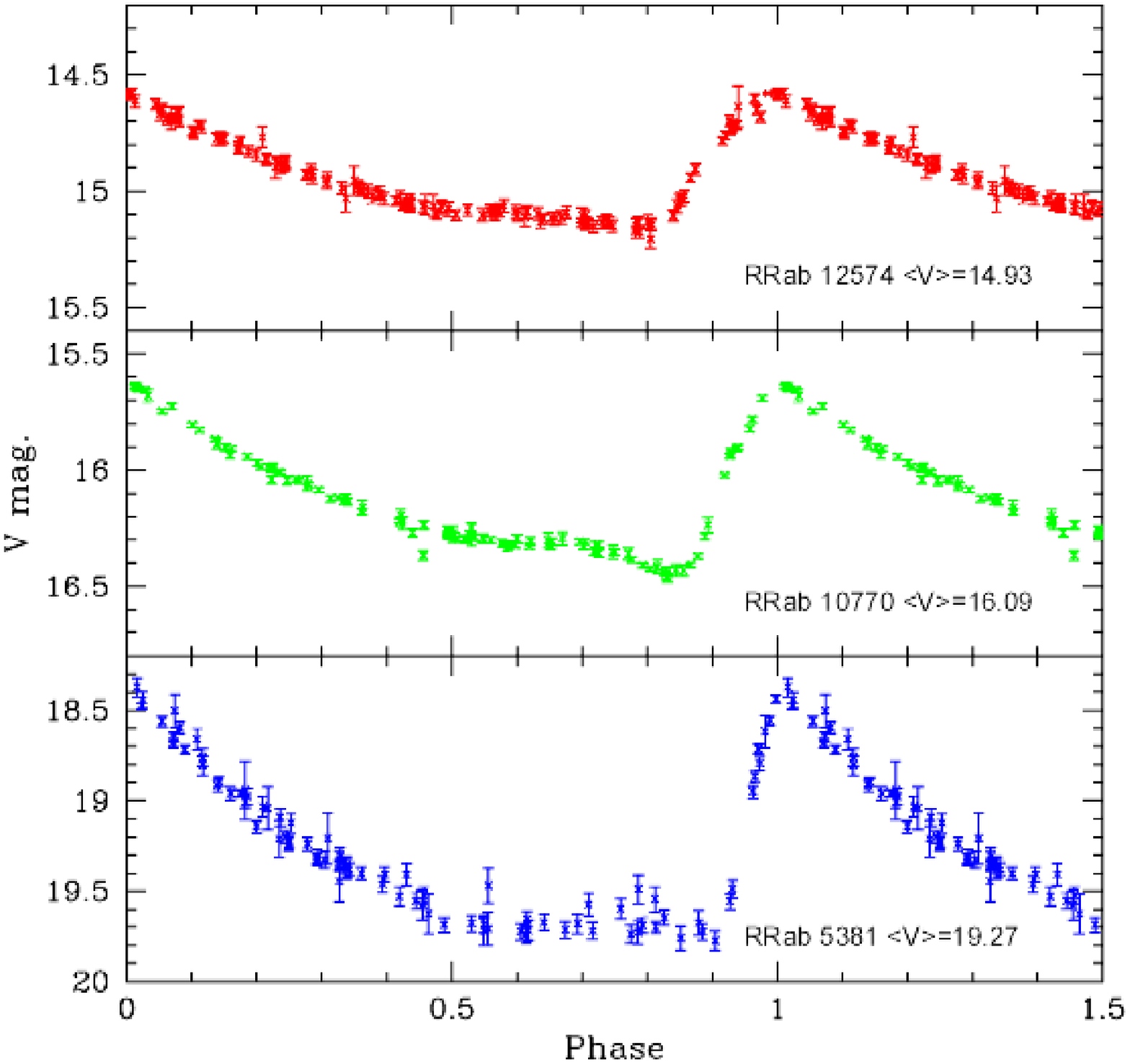}}
     \subfigure[Relative Qst*-band brightness versus orbital phase $\theta$, from LSQ observations.]{\includegraphics[width=2.0in]{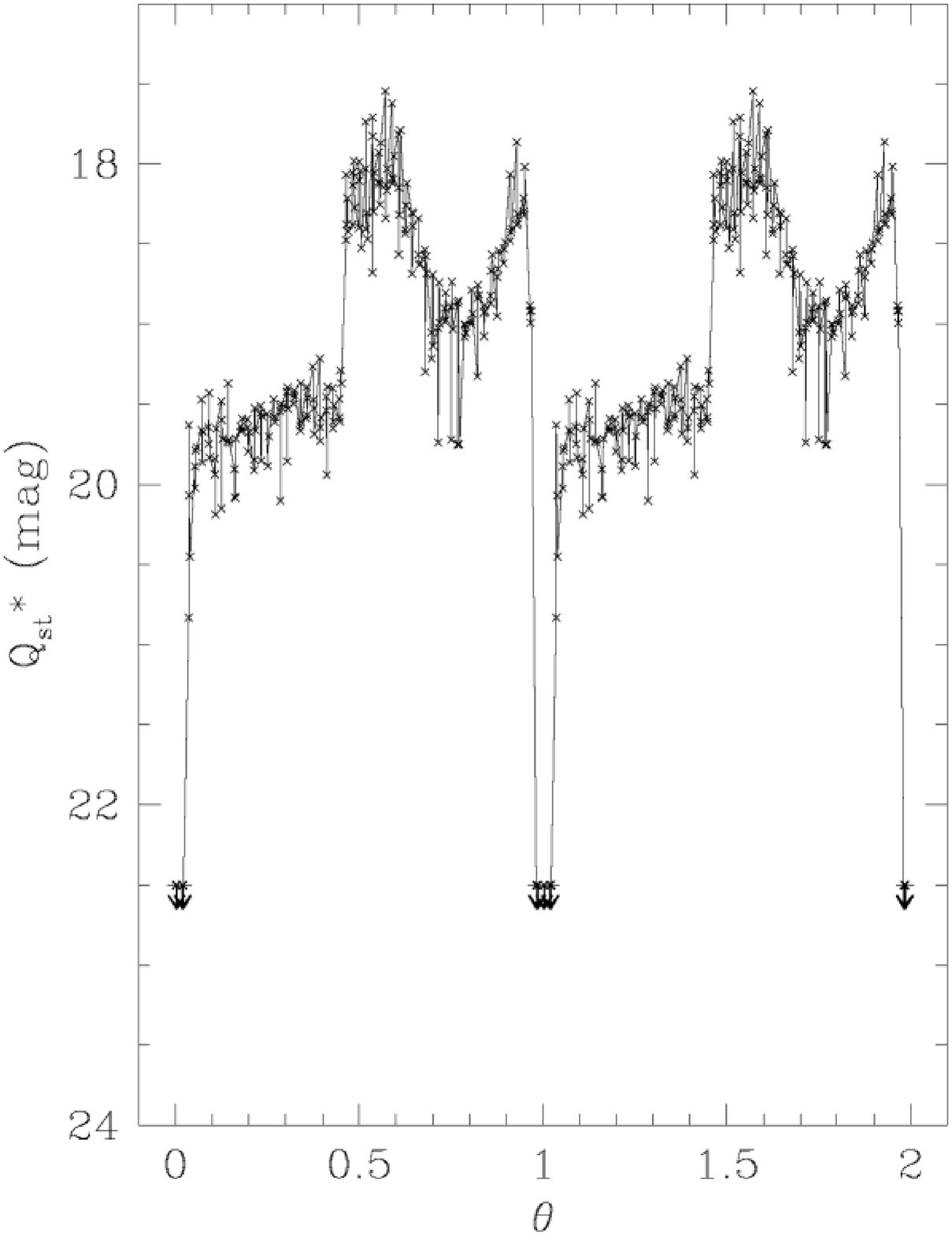}}
   \end{tabular}
 \end{center}
 \caption{LSQ Example Transient Detections.}\label{fig5}
\end{figure}

The LSQ RR Lyrae star (RRLS) survey is searching the galactic halo for RRLS that have V magnitudes between roughly 14 and 20.  Because RRLS are excellent standard candles, they provide a powerful probe of the density distribution of the halo, which is being examined for halo substructure. Plotted are the V magnitude light-curves for three type ab RRLS (Fig.\,\ref{fig5}(a)), which illustrate the typical photometric precisions at these magnitudes.  From their mean V magnitudes, we estimate that RRab 12574, 10770, and 5381 lie 7, 13, and 52 kpc from the Sun, respectively.

\subsection{A Deep Eclipsing CV}

A deep eclipsing cataclysmic variable (\cite[Rabinowitz et al. 2011a]{Rabinovitz_etal11a}) was discovered with eclipse depths $>$5.7
magnitudes, orbital period 94.657 min, and peak brightness V$\sim$18 at J2000 position 17h 25m
54.8s, -64 deg 38 min 39 sec. Light curves in B, V, R, I, z and J were obtained with SMARTS 1.3-m and 1.0-m
telescopes at Cerro Tololo and spectra from 3500 to 9000{\AA} with the SOAR 4.3-m telescope at Cerro Pachon.
The optical light curves (Fig.\,\ref{fig5}(b)) show a deep, 5-min eclipse immediately followed by a
shallow 38-min eclipse and then sinusoidal variation. No eclipses appear in J. During the deep eclipse
the measure of V-J $>$ 7.1 corresponds to a spectral type M8 or later secondary. The spectra show strong
Hydrogen emission lines, Doppler broadened by 600 - 1300 km s-1, oscillating with radial velocity that
peaks at mid deep eclipse with semiamplitude 500 $\pm$ 22 km s-1. It is suggested that LSQ172554.8-643839
is a polar with a low-mass secondary viewed at high inclination.


\subsection{A Dwarf Novae}

\begin{figure}[h]
 \begin{center}
   \begin{tabular}{c}
     \subfigure[Before and after discovery images of the dwarf novae.]{\includegraphics[width=1.5in]{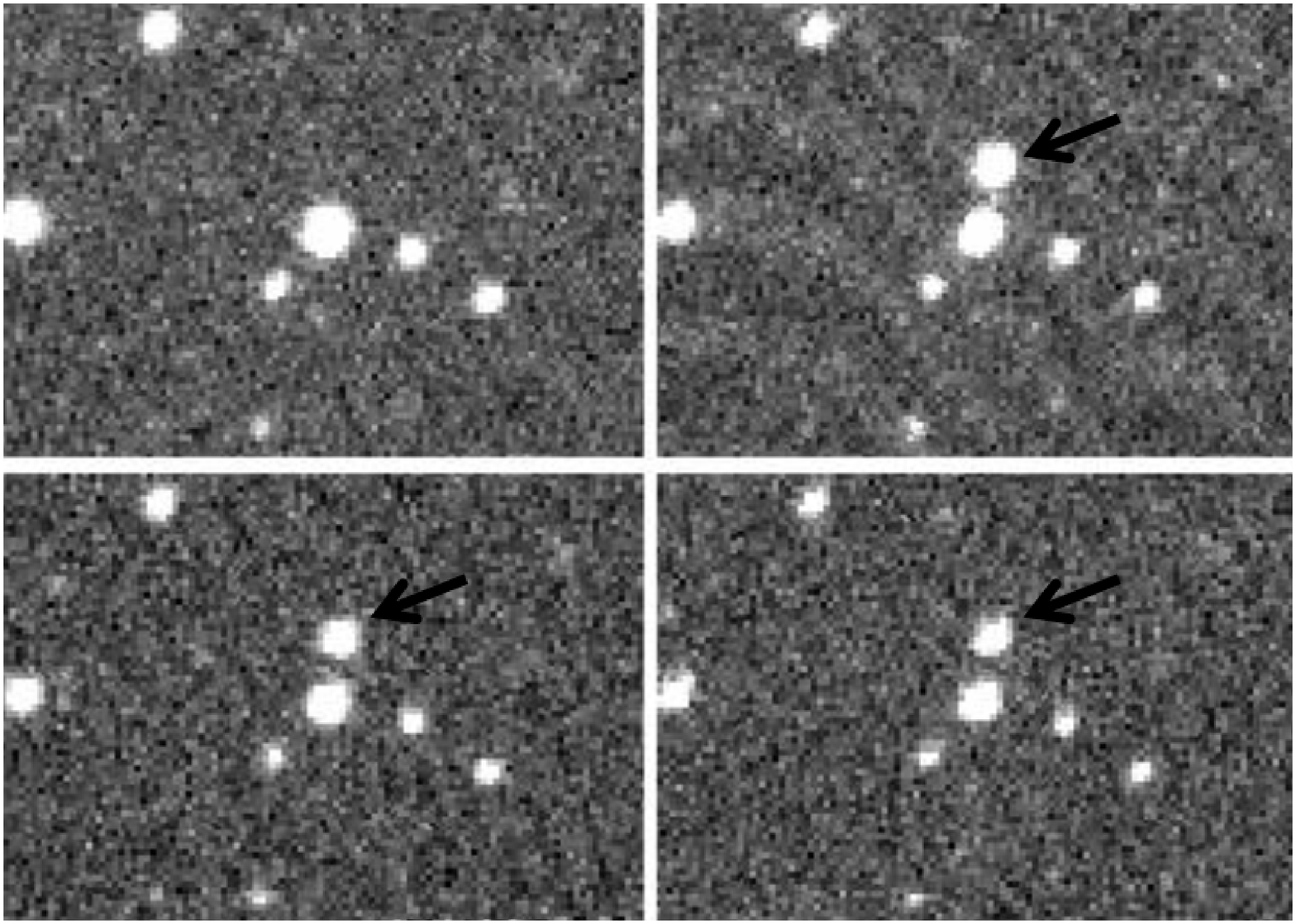}}
     \subfigure[Relative Qst* light curve of LSQ J16531857-1617542, upper limits and error barred detections.]{\includegraphics[width=1.9in]{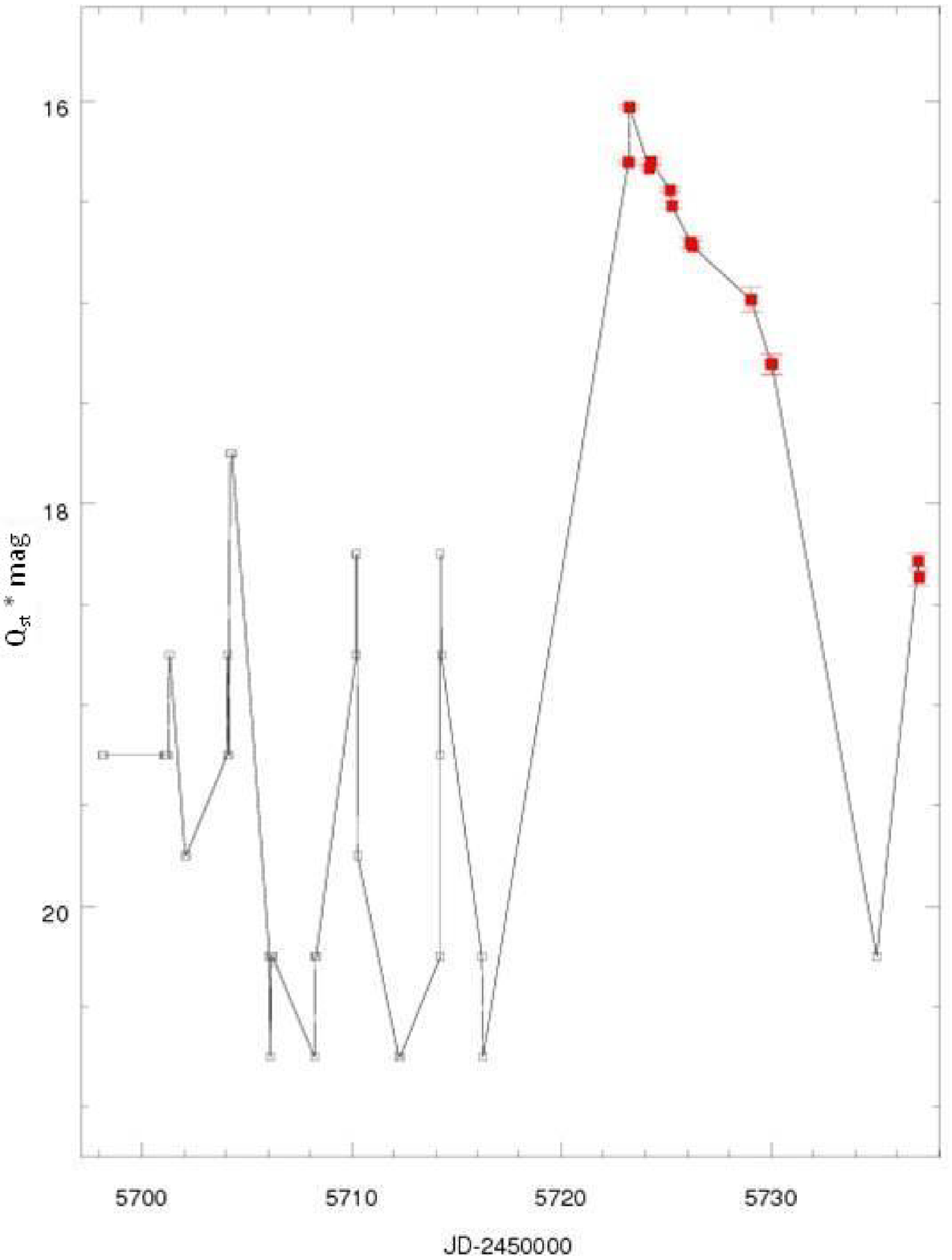}}
   \end{tabular}
 \end{center}
 \caption{The first followed up LSQ subtraction pipeline transient candidate.}\label{fig6}
\end{figure}

An apparent dwarf nova was discovered (\cite[Rabinowitz et al. 2011b]{Rabinovitz_etal11b}) on June 11.046 (when the variable was at magnitude R = 16.3) and
June 11.063 UT (at R = 16.0)(Fig.\,\ref{fig6}(a) and (b)). A faint source is reported at this position in the Guide Star Catalog V2.3.2
(with Bj = 20.76). Simultaneous visible and J-band observations were taken with ANDICAM on the 1.3-m SMARTS
telescope at Cerro Tololo. A spectrum
(range 0.350-0.966 nm) taken on June 26 with GMOS on the Gemini South telescope reveals strong H-alpha and
H-beta emission lines, with H-alpha clearly double-peaked, indicating the presence of an accretion disk
with rotational velocity 1000 km/s.

\smallskip\par
\smallskip\par
\smallskip\par

This work has been made possible with the help of NERSC computer resources as well as funding from DOE and NASA.


\begin{thebibliography}{}
\bibitem[Andrews et al. (2008)]{Andrews_etal08}
{Andrews, P., Baltay, C., Bauer, A., Ellman, N., Jerke, J., Lauer, R., Rabinowitz, D., Silge, J.} 2008,
\textit{PASP}, 120, 703A

\bibitem[Baltay et al. (2007)]{Baltay_etal07}
{Baltay, C., Rabinowitz, D., Andrews, P., Bauer, A., Ellman, N., Emmet, W., Hudson, R., Hurteau, T., Jerke, J., Lauer, R., Silge, J., Szymkowiak, A.} 2007,
\textit{PASP}, 119, 1278B

\bibitem[Rabinowitz et al. (2011a)]{Rabinowitz_etal11a}
{Rabinowitz, D., Tourtellotte, S., Rojo, P., Hoyer, S., Folatelli, G., Coppi, P., Baltay, C., Bailyn, C.} 2011,
\textit{ApJ}, 732, 51

\bibitem[Rabinowitz et al. (2011b)]{Rabinowitz_etal11b}
{Rabinowitz, D., Baltay, C., Ellman, N., Hadjiyska, E., Tourtellote, S.} 2011,
\textit{CBET}, 2757, 1R
\end{thebibliography}
\end{document}